\documentclass[journal, twocolumn, 10pt, letter]{IEEEtran}

\usepackage[margin=0.7in]{geometry}

\usepackage{amsmath,cite}
\usepackage{amssymb}
\usepackage{color,epsfig}
\usepackage{cite}

\newtheorem{theorem}{Theorem}

\newtheorem{remark}{Remark}
\newcommand{\remarkend}{ \IEEEQEDopen}

\newenvironment{myalign}{\allowdisplaybreaks \align}{\endalign}

\newcommand{\cC}{{\mathcal C}}
\newcommand{\xvec}{{\mathbf x}}
\newcommand{\td}{{\tilde d}}

\begin{document}

\title{MIMO Broadcast Channel with an Unknown Eavesdropper: Secrecy Degrees of Freedom}

\author{ \IEEEauthorblockN{Xiang He\IEEEauthorrefmark{1}, Ashish
    Khisti\IEEEauthorrefmark{2}, Aylin Yener\IEEEauthorrefmark{1}} \\
  \IEEEauthorblockA{\IEEEauthorrefmark{1} Electrical Engineering
    Department, The Pennsylvania
    State University, University Park, PA 16802\\
    \IEEEauthorrefmark{2}Dept. of Electrical and Computer Engineering,
    University of Toronto,  Toronto, ON, M5S 3G4, Canada\\
    \textit{hexiang@ieee.org, akhisti@comm.utoronto.ca, yener@ee.psu.edu}}\thanks{A portion of this paper was presented in~\cite{xiang2011globecom}. A.~Khisti's work was supported by a NSERC Discovery Grant and the Qatar National Research Fund, NPRP Award. A.Yener's work was supported in part by NSF Grants 0964362 and 1319338}}
\maketitle

\vspace{-2em}

\begin{abstract}
We study a multi-antenna broadcast channel with two legitimate receivers and an external eavesdropper.
We assume that the channel matrix of the eavesdropper is unknown to the legitimate terminals 
but satisfies a maximum rank constraint. As our main result we characterize the associated secrecy degrees of freedom for the broadcast channel with common and private messages. 
We show that a direct extension of the single-user wiretap codebook does not achieve the 
 secrecy degrees of freedom. Our proposed optimal scheme involves decomposing the signal space 
into a common subspace, which can be observed by both receivers, and private subspaces which can be observed by only 
one  of the receivers, and carefully transmitting  a subset of messages in each subspace.
We also consider the case when each user's private
message must additionally remain confidential from the other legitimate receiver and characterize the s.d.o.f.\
region in this case.
\end{abstract}

\section{Introduction}
\label{sec:introduction}
Claude Shannon~\cite{shannonsecrecy} pioneered the information theoretic approach
for secure communication.  Shannon's notion of {\em perfect secrecy} requires that the information message 
and the eavesdropper's observation be statistically independent. 
This framework was later extended to different network
models, see e.g.,~\cite{csiszar1978bcc, ender-special, liangcompound, lai2006rec, ekrem-2009,Ebrahimthesis}, where
various relaxations of perfect secrecy were  considered and the associated secrecy capacity was studied.
 In recent years there has been a growing interest in using multiple antennas for securing wireless
networks, see e.g.,~\cite{khisti:stm, khisti:stm2, liu2010multiple, kobayashi-2009,ulukus-1,ulukus-2,khisti-zhang}. 
In these works generally some sort of side information of the eavesdropper's channel --- either complete, partial or statistical ---  is made available to the legitimate terminals.  In contrast reference~\cite{xiang2010arb} considers a single-user Gaussian MIMO wiretap
channel when the eavesdropper's channel is unknown and  time-varying,
but satisfies a maximum-rank constraint. The existence of a coding scheme that
simultaneously attains strong secrecy  against all feasible  eavesdropper channels is  
established. 
Furthermore, two receiver broadcast and multiple-access channels (MAC) are also treated in~\cite{xiang2010arb} when each of the legitimate terminals has an equal number of antennas and the optimality of a time-sharing based scheme is established in either case. Recently a complete characterization of the secure degrees of freedom for the two-user MIMO MAC channel with an arbitrarily varying eavesdropper has been obtained in~\cite{mimomac}.


In this paper, we consider the two-receiver MIMO broadcast channel when there is a private
message for each receiver as well as a common message for both receivers. The messages must remain confidential
from  an eavesdropper. We assume that the channel matrices of the legitimate terminals are known 
to all the terminals whereas the channel matrix of the eavesdropper is only known to the eavesdropper.
However an upper bound on the rank of the eavesdropper channel matrix, or equivalently the
maximum number of antennas at the eavesdropper is known.
We characterize the secrecy degrees of freedom (s.d.o.f.)
region for such a model, as well as a variation when the private messages must also remain mutually confidential from the other receiver. 
Interestingly the optimal scheme does not follow from a direct extension of the techniques used in the single-user channel
\cite{xiang2010arb}. Such an approach introduces  independent
randomization in each user's codebook and creates higher than necessary
interference between users.  Instead our proposed approach involves decomposing the
signal space into a common subspace seen by both receivers and private subspaces
seen by only one of the receivers; and transmitting a fictitious message of just
enough rate such that it can simultaneously provide secrecy for \textit{both} users.
We show that the s.d.o.f.\ achieved by the proposed scheme are in-fact optimal and meet
the natural cut-set upper bound for the broadcast network. In contrast the scheme based on 
the  single-user codebooks is sub-optimal in general.
We limit our work to the case when the eavesdropper's channel is fixed throughout
the duration of communication, but unknown to the legitimate terminals.

We note that the literature on secure network coding~\cite{cai2002snc, silva2008universal} is also related to our setup. 
The most closely related paper to our present work is reference~\cite{ashish2010secureisit},
which considers an extension of secure network coding  for broadcasting to two receivers.
The combined message of both  users maps to a syndrome vector of a maximum rank distance (MRD) code (c.f.~\cite{silva2008universal}).  
The parity-check matrix of the MRD code is designed to be in a certain systematic form,  so that
each receiver is able to recover the desired message from the observed sequence. 
While the results in the present paper are structurally similar to~\cite{ashish2010secureisit},
 our underlying approach is very different. Instead of attempting a direct extension of the MRD codes to Gaussian channels we propose a random coding technique where an explicit fictitious message, shared by both the receivers is also transmitted.
The key insight in our proposed scheme is to minimize the rate associated with this fictitious  messages by using a 
carefully constructed signal space decomposition.

In the remainder of the paper, we present the system model in
Section~\ref{sec:model} and a summary of the main results in
Section~\ref{sec:results}.  In section~\ref{sec:GSVD} we present
a reduction of the MIMO broadcast
channel into independent parallel channels, which is based on the
Generalized Singular Value Decomposition. Thereafter sections~\ref{sec:thm1}
and~\ref{sec:thm2} provide proofs of the main results and section~\ref{sec:conclusion}
concludes the paper.

Throughout this paper we only focus on the case of two legitimate receivers. 
Unfortunately an  extension of our results to more than two receivers may not be straightforward. 
Indeed to the best of our knowledge, the degrees of freedom of the MIMO broadcast
channel even without secrecy constraints remains an open problem when both common and private (individual)
messages are considered. The well known compound MIMO broadcast channel is a special case of this setup~\cite{compound-1,compound-2,compound-3}. Furthermore our lower bound involves the GSVD transform, whose direct extension to more than two channels does not appear straightforward and therefore we only limit to the case of two legitimate receivers. Nevertheless we believe that the setup considered in this paper is of practical significance. 
Further note that in this paper we only consider the secrecy degrees of freedom (s.d.o.f.), which measures the \emph{pre-log} of the achievable rates. While a considerably coarse measure of the capacity region, the s.d.o.f.\ analysis is tractable and provides important insights into the optimal scheme in the high signal-to-noise-ratio regime.
For some prior works on s.d.o.f., see e.g.,~\cite{liangcompound,yang2011secrecy,xiang2010arb, mimomac, ulukus-1, ulukus-2}.


\section{System Model}
\label{sec:model}
\begin{figure}[t]
  \centering 
\includegraphics[scale=0.5]{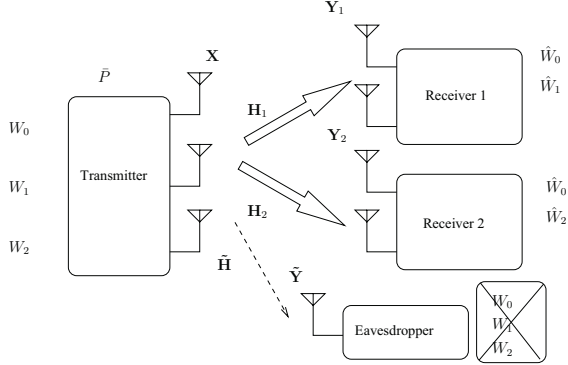}
  \caption{The MIMO Broadcast Wiretap Channel where $N_T=3,
    N_{R_1}=N_{R_2}=2, N_E=1$.}
  \label{fig:mimobc}
\end{figure}

We consider a MIMO Broadcast (BC) wiretap channel with two receivers,
as shown in Figure~\ref{fig:mimobc}. We assume that the number of antennas
at the transmitter, receiver $1$, receiver $2$ and the eavesdropper are given by
$N_T,$ $N_{R_1},$ $N_{R_2}$ and $N_E$ respectively:
\begin{myalign}
& \mathbf{Y}_t(i) = {\mathbf{H}_t \cdot \mathbf{X} }(i)  + \mathbf{Z}_t(i), \quad t=1,2 \label{eq:mainchannel} \\ 
& \mathbf{\tilde Y}(i) =  {\mathbf{\tilde H}\cdot \mathbf{X}(i) }  \label{eq:eve-model}
\end{myalign}
where $\mathbf{Y}_t(i)$ denotes the symbols received at the
legitimate receivers at time $i$ whereas  $\mathbf{\tilde Y}(i)$ denotes the received
symbols at the eavesdropper,  $\mathbf{H}_t$ and $\mathbf{\tilde
  H}$ are the channel matrices and $\mathbf{Z}_t$ is the
additive Gaussian noise observed by the intended receiver $t$, which
is composed of independent rotationally invariant complex Gaussian
random variables with unit variance.

\begin{remark}
Note that in~\eqref{eq:eve-model} 
we do not assume any noise on the eavesdropper's channel.  In practice
the eavesdropper's channel will have some additive noise and its observation will be
a degraded version of~\eqref{eq:eve-model}.  Thus our achievability results immediately
apply to such degraded channels. As such the model we study in~\eqref{eq:eve-model}
is the worst case model among all eavesdropper channels. 
While the converse for the above model does not directly apply,
it can be easily extended to show  that the s.d.o.f.\ region does not increase when the eavesdropper's channel has additive noise.
\end{remark}

Note that the rank of $\mathbf{\tilde H}$ in~\eqref{eq:eve-model} is upper bounded by
$N_E,$ which is known to all the terminals.  The realization $\mathbf{\tilde H} = {\mathbf{\tilde{h}}}$
is  revealed only to the eavesdropper and not to the legitimate terminals.  
Throughout this paper we will assume that $N_E < N_T$, since otherwise the secrecy degrees of freedom is zero
in the single-user setup~\cite{xiang2010arb}. Similarly if either $N_E \ge N_{R_1}$ or $N_E \ge N_{R_2}$ the s.d.o.f.\
for at least one of the receivers is zero and the problem degenerates to the single user case. Thus we also assume that $N_E < \min(N_{R_1}, N_{R_2})$.
Our proposed setup guarantees confidentiality regardless
  of the particular channel realization of the eavesdropper. In contrast, the channel matrices $\mathbf{H}_t$ are
known to both the legitimate parties and the eavesdropper(s).

In practice note that the complete lack of the eavesdropper's CSI at the legitimate terminals is far more realistic than the previous models studied with partial or full knowledge, given that the eavesdropper is a passive observer who does not transmit signals. The limit on $N_E$ can be justified since the passive eavesdropping device must be stealth, and hence be  limited in number of antennas due to size limitations.

The input symbols in~\eqref{eq:eve-model}, denoted by ${\mathbf{X}}(i),$ must satisfy the average power constraint:\begin{myalign}
E\left[ \frac{1}{ n} {\sum_{i=1}^{{n}}}  \mathrm{trace}\left({\mathbf{X}}(i)
  {\mathbf{X}}^H(i)\right) \right]\le \bar P. \label{eq:powerbc}
\end{myalign}

We next define the associated secure broadcast code.
Receiver $t$ must decode a  confidential message $W_t$, and a common
confidential message $W_0$ over $ n$ channel
uses. The messages $(W_0,W_1,W_2)$ must be kept jointly confidential from the
eavesdropper.  Let $\mathbf{\tilde Y}_{\mathbf{\tilde h}}^n$ denote the signals
received by the eavesdropper when its channel matrix 
$\mathbf{\tilde H}$ equals $\mathbf{\tilde h}$. We impose the following secrecy constraint:
\begin{myalign}
  w\bigg( \mathop {\lim }\limits_{ n \to \infty } \frac{1}{{n}} \sup_{\mathbf{\tilde h}} I ({W_0, W_1 ,W_2
      ;  \mathbf{\tilde Y}_{\mathbf{\tilde h}}^n } )\bigg) = 0, \label{eq:secrecybc}
\end{myalign}
where $w(x) = \lim_{\bar P \rightarrow \infty}\frac{x}{\log_2 \bar
  P}$.  To interpret the secrecy constraint in~\eqref{eq:secrecybc},
  note that  $\frac{1}{{n}} I ({W_0, W_1 ,W_2
      ;  \mathbf{\tilde Y}_{\mathbf{\tilde h}}^n } ) $
is the information leakage-rate~\cite[sec 22.1, pp.~550]{elGamal} at the eavesdropper.
The constraint in~\eqref{eq:secrecybc}  only requires that the pre-log of the asymptotic leakage-rate
at the eavesdropper be zero. Note that this condition is weaker than the usual notion of {\em weak secrecy}
which requires that the  information leakage-rate approach must zero asymptotically in $n$. 
Strictly speaking, we should refer to~\eqref{eq:secrecybc}
as the secrecy-DOF constraint, but we drop the ``DOF" for simplicity in this paper.
We primarily consider this notion as it suffices to highlight the key ideas in the 
coding scheme proposed in the paper. We point the reader to~\cite{xiang2010arb} for the
for the analysis of strong secrecy and time-varying eavesdropper channels in the single user case.

When an additional constraint of mutual privacy is imposed on the messages $W_1$ and $W_2$
we further require that:
\begin{align}
&w\bigg(\lim_{{ n}\rightarrow \infty} \frac{1}{ n}I(W_1; {\mathbf{Y}}_2^{ n})\bigg)=0 \label{eq:mut-priv-1}\\
&w\bigg(\lim_{{ n}\rightarrow \infty} \frac{1}{ n}I(W_2; {\mathbf{Y}}_1^{ n})\bigg)=0\label{eq:mut-priv-2}
\end{align}
The secrecy rate tuple $(R_{s,0}, R_{s,1}, R_{s,2})$ is achievable if 
$R_{s,i}=\lim_{ n \to \infty} \frac{1}{ n} H(W_i),
i=0,1,2 \label{eq:def_R_e_k}$ and a sequence of encoding and decoding
functions exists (indexed by $ n$) such that the error probability in decoding of $\{W_0, W_t\}$ by receiver $t$
approaches zero as $ n\rightarrow \infty$ and furthermore the secrecy constraints~\eqref{eq:secrecybc} is
satisfied. In addition when a mutual privacy constraint is imposed we also require that~\eqref{eq:mut-priv-1}
and~\eqref{eq:mut-priv-2} be satisfied.

In this paper, we use the secrecy degrees of freedom (s.d.o.f.) region
as a characterization of the high SNR behaviour of the secrecy capacity
for this channel. The s.d.o.f. pair $(d_0, d_1,d_2)$ is achievable if there exists a sequence of achievable rates
$(R_{s,0}({\bar P}), R_{s,1}({\bar P}), R_{s,2}({\bar P}))$, indexed by ${\bar P}$, such that $d_i=\limsup_{\bar P \to \infty}\frac{ R_{s,i}({\bar P})}{\log_2 \bar
    P}$ for $i=0,1,2$. The set of all achievable $(d_0, d_1, d_2)$ is called the s.d.o.f.\ region.

\section{Main Results}
\label{sec:results}

The secrecy degrees of freedom region  is characterized using  rank of the associated channel matrices.
Let $r_1, r_2$ be the rank of $\mathbf{H}_1$ and $\mathbf{H}_2$ respectively. Let 
  \begin{align}
  r_0 = \mathrm{rank}\left[ \begin{array}{c} {\mathbf{H}_1} \\ {\mathbf{H}_2}\end{array}\right] 
  \label{def-rank}
  \end{align}
and let
\begin{myalign}
  s = r_1 + r_2 - r_0\label{def:s}
\end{myalign}
be the dimension of the common row-space of $\mathbf{H}_1$ and $\mathbf{H}_2$.

\begin{theorem}
  The secrecy degrees of freedom region for the MIMO broadcast wiretap
  channel in absence of the mutual privacy constraint is given by all non-negative triples $(d_0, d_1, d_2)$ that satisfy the following constraints:
\begin{myalign}
&  0 \le d_0 + d_i \le \{r_i - N_E\}^+, \quad i=1,2 \label{eq:bnd-t1}\\
&  0 \le d_0 + d_1 + d_2 \le \{ r_0 - N_E\}^+ \label{eq:sumratebound}
\end{myalign}
where we use the notation that $\{v\}^+ \stackrel{\Delta}{=} \max(0,v)$. \remarkend
\label{thm:mimobcsdof}
\end{theorem}

The inequalities in Theorem~\ref{thm:mimobcsdof} can be interpreted as the cut-set bounds in the broadcast network. The two inequalities in~\eqref{eq:bnd-t1} are single user bounds, whereas the inequality in~\eqref{eq:sumratebound} corresponds to the case when both the receivers are allowed to cooperate. Our proof of the coding
theorem shows that these bounds are also achievable, whereas the converse involves selecting the specific 
eavesdropper channel gains that lead to these upper bounds.

\begin{theorem}
  The secrecy degrees of freedom region for the MIMO broadcast wiretap
  channel in presence of the mutual privacy constraint~\eqref{eq:mut-priv-1} and~\eqref{eq:mut-priv-2} 
  consists of all non-negative triples $(d_0, d_1, d_2)$ that satisfy the following constraints:
\begin{myalign}
&  0 \le d_0 + d_i \le \{r_i - N_E\}^+,\quad  i=1,2, \label{eq:bnd-t2-1}\\
& 0 \le d_i \le \{r_i - s\}^+,\quad  i=1,2. \label{eq:bnd-t2-2}
\end{myalign}\remarkend
\label{thm:mimobcsdof-2}
\end{theorem}

The inequalities in~\eqref{eq:bnd-t2-1} corresponds to single-user bounds associated with each receiver, whereas the inequalities in~\eqref{eq:bnd-t2-2} correspond to transmission of only a private message to each receiver, with the other
receiver as the only eavesdropper in the network. Note that the sum-constraint is not active in Theorem~\ref{thm:mimobcsdof-2}.

\begin{figure}
\includegraphics[width=\columnwidth]{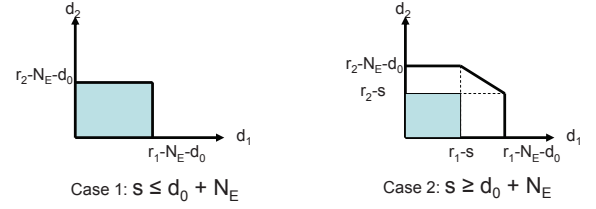}

\caption{Achievable Secrecy Degrees of Freedom for the two-user MIMO broadcast channel with an external eavesdropper. We fix the
s.d.o.f. of the common message to $d_0$ and plot $(d_1, d_2)$. The shaded area corresponds the s.d.o.f. achievable
with mutual privacy constraint. The figure on the left corresponds to the case when $s \le d_0 + N_E$ while the figure on the right corresponds to the case when $s \ge d_0 + N_E$.}
\label{fig:dim}
\end{figure}

Fig.~\ref{fig:dim} compares the results in Theorem~\ref{thm:mimobcsdof} and~\ref{thm:mimobcsdof-2}.
We observe that the structure of the capacity region takes one of two forms.
In case $1$, we assume that
\begin{myalign}
 N_E' \stackrel{\Delta}{=} d_0 + N_E \ge s.\label{eq:inthiscase}
\end{myalign}
It can be  verified that the  two constraints in~\eqref{eq:bnd-t1}
imply~\eqref{eq:sumratebound} (by adding the constraints~\eqref{eq:bnd-t1} and using ~\eqref{def:s} and~\eqref{eq:inthiscase}).
Therefore the projection of the s.d.o.f.\ region in the $(d_1, d_2)$ plane reduces to a rectangle
\begin{align}
d_i \le \{r_i -N_E'\}^+, \quad i=1,2
\end{align}
and the sum-rate constraint is not active.
Furthermore upon examining
\eqref{eq:bnd-t2-1} and \eqref{eq:bnd-t2-2}, one can conclude that the same region is also achieved in
Theorem~\ref{thm:mimobcsdof-2}, where an additional mutual privacy constraint is imposed. 

In case $2$, which corresponds to $N_E' \le s,$ 
 the sum-rate constraint~\eqref{eq:sumratebound} is active.
It can be easily verified that the constraints
\eqref{eq:bnd-t1} and~\eqref{eq:sumratebound}, using~\eqref{def:s}, reduce to
$d_i \le r_i - N_E'$, and $d_1 + d_2 \le r_1 + r_2 - s - N_E'$.
Thus as shown in Fig.~\ref{fig:dim},  $(d_1, d_2) = (r_1-s, r_2 - N_E')$ and $(d_1, d_2)= (r_1-N_E',
r_2-s)$ are the two corner points in the $(d_1, d_2)$ plane. 
Furthermore examining~\eqref{eq:bnd-t2-1},~\eqref{eq:bnd-t2-2}
in Theorem~\ref{thm:mimobcsdof-2},  the active constraints in the case when 
$N_E' \le s$ are $d_i \le \{r_i -s\}^+$ for $i=1,2$.
This region is in general  smaller than the region achieved in Theorem~\ref{thm:mimobcsdof}. 


As a final remark we note that when $N_E=0$, i.e., the eavesdropper is absent, the result here is equivalent to
the degrees of freedom  for the two-user MIMO broadcast channel with common and private messages~\cite{ersen2010globecom}.

\section{Generalized Singular Value Decomposition}
\label{sec:GSVD}

\begin{figure}
\begin{center}
\includegraphics[scale=0.5]{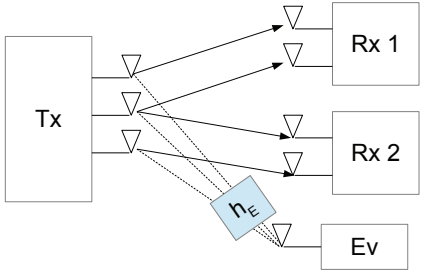}
\caption{The decomposition of the $3\times 2 \times 2 \times 1$ MIMO Broadcast Wiretap channel (cf.~Fig.~\ref{fig:mimobc}) using the GSVD transform. The channel matrices of the legitimate receivers are scaled versions of $[{\bf I}_2, {\bf 0}]$ and $[{\bf 0}, {\bf I}_2]$ respectively, while the eavesdropper channel matrix is of rank at most $1$.}
\end{center}
\end{figure}

A common element in code construction in both Theorem~\ref{thm:mimobcsdof}
and~\ref{thm:mimobcsdof-2} is the Generalized Singular Value Decomposition (GSVD)~\cite{paige1981towards}
previously used in \cite{khisti:stm2 } in the MIMO 
wiretap channel literature. The GSVD transform can be used to decompose the channel in \eqref{eq:mainchannel} 
into parallel and independent channels, which are more amenable to
analysis. 
\begin{theorem}
\label{thm:GSVD}
\cite{paige1981towards} There exist unitary matrices $\mathbf{U},
\mathbf{V}, \mathbf{W}, \mathbf{Q}$ and a nonsingular upper triangular
matrix $\mathbf{R}$ such that
  \begin{myalign}
& \mathbf{U}^H \mathbf{H}_1 \mathbf{Q} = \mathbf{\Sigma}_{1 (N_{R_1}
  \times r_0)} \left[ {\mathbf{W}^H \mathbf{R}_{(r_0 \times r_0)},\mathbf{0}}
\right]_{(r_0 \times N_T)} \label{eq:H1_decomp}\\ 
& \mathbf{V}^H \mathbf{H}_2 \mathbf{Q} = \mathbf{\Sigma}_{2 (N_{R_2}
  \times r_0)} \left[ {\mathbf{W}^H \mathbf{R}_{(r_0 \times r_0)},\mathbf{0}}
\right]_{(r_0 \times N_T)} \label{eq:H2_decomp}\\ 
& \mathbf{\Sigma}_1  = \left[ {\begin{array}{*{20}c}
   {\mathbf{I}_{1 (\tilde r_1 \times \tilde r_1)} } & {} & {}  \\
   {} & {\mathbf{S}_{1 (s \times s)} } & {}  \\
   {} & {} & {\mathbf{O}_{1( (N_{R_1}-\tilde r_1 -s)\times \tilde r_2)} }  \\
\end{array}} \right] \label{eq:Sigma_1}\\
&\mathbf{\Sigma}_2  = \left[ {\begin{array}{*{20}c}
   {\mathbf{O}_{2 ((N_{R_2}- \tilde r_2-s) \times \tilde r_1)}} & {} & {}  \\
   {} & {\mathbf{S}_{2 (s \times s)} } & {}  \\
   {} & {} & {\mathbf{I}_{2 ( \tilde r_2 \times \tilde r_2) } }  \\
\end{array}} \right] \label{eq:Sigma_2}
   \end{myalign}    
   where $ {\mathbf{S}_i }, i=1,2$ are $s \times s$ diagonal matrices
   with positive real elements on the diagonal line, $\mathbf{I}_i,
   i=1,2$ are $\tilde r_i \times \tilde r_i$ identity matrices 
   and   the matrices $\mathbf{O}_i, i=1,2$ are zero
   matrices. For clarity, the dimension of each matrix is shown in the
   parenthesis in the subscript.
Recall from~\eqref{def-rank} and~\eqref{def:s} that $r_0$ 
  equals the rank of $
\left[
\begin{array}{c}
{\mathbf H}_1  \\
{\mathbf H}_2   
\end{array}
\right]$, $r_1$ and $r_2$
  equal the rank of ${\mathbf{H}_1}$ and ${\mathbf H}_2$,  
  and  we let $s = r_1 + r_2-r_0$.
The constants $\tilde r_i,$  for $i=1,2,$ are given by $\tilde{r}_i = r_i-s$.
 \remarkend
\end{theorem}

We next demonstrate the simultaneous reduction of the channel matrices $\mathbf{H}_i, i=1,2$
into parallel and independent channels using the GSVD transform. Let the decomposition of ${\mathbf H}_i$
be as in~\eqref{eq:H1_decomp} and~\eqref{eq:H2_decomp}. 
We left-multiply the transmitted signals with $\mathbf{Q}$, left-multiply the received signals with
$\mathbf{U}^H$ at receiver $1$, and left-multiply the received signals
with $\mathbf{V}^H$ at receiver $2$. Since $\mathbf{Q}$, $\mathbf{U}^H$
and $\mathbf{V}^H$ are all unitary matrices the setup is equivalent to the following:
\begin{myalign}
  & \mathbf{Y}_t(i) = \mathbf{\Sigma}_{t} \left[ {\mathbf{P}_{(r_0 \times r_0)},\mathbf{0}}\right] \mathbf{X} (i)  + \mathbf{Z}_t(i), \quad t=1,2,  \\ 
& \mathbf{\tilde Y}(i) =  {\mathbf{\tilde H} \mathbf{X}(i) },  
\end{myalign}
where we have introduced the matrix ${\mathbf{P}} \stackrel{\Delta}{=} {\mathbf{W}}^H {{\mathbf R}}$.  We also set the last $N_T-r_0$ component of $\mathbf{X}$
to zero and design the achievable scheme for the following channel model:
\begin{myalign}
  & \mathbf{Y}_t(i) = \mathbf{\Sigma}_{t} \mathbf{P}_{(r_0 \times r_0)} \mathbf{X}_{(r_0 \times 1)} (i)  + \mathbf{Z}_t(i), t=1,2,  \\
  & \mathbf{\tilde Y}(i) = \{{\mathbf{\tilde H}{\mathbf{Q}}\}_{(N_E \times r_0)} \mathbf{X}_{(r_0 \times 1)}(i) },
\end{myalign}
where
$\{{\mathbf{\tilde H}{\mathbf{Q}}\}_{(N_E \times r_0)}}$ denotes the first $r_0$
columns of the matrix $\mathbf{\tilde H}{\mathbf{Q}}$.  Since $\mathbf{P}$ is nonsingular,
without loss of generality, we can view $\mathbf{P} \mathbf{X}_{(r_0
  \times 1)} (i)$ as the input to the channel. The main channel can
then be expressed as $ \mathbf{Y}_t(i) = \mathbf{\Sigma}_{t}
\mathbf{X}_{(r_0 \times 1)} (i) + \mathbf{Z}_t(i), t=1,2$ and the
eavesdropper channel reduces to $\mathbf{\tilde Y}(i) =
\{{\mathbf{\tilde H}{\mathbf{Q}}\}_{(N_E \times r_0)} \mathbf{P}^{-1}\mathbf{X}_{(r_0
    \times 1)}(i) } $. Note that the eavesdropper channel state matrix
is arbitrary, and ${\mathbf Q}$ is a unitary matrix, and thus it can be easily seen that the rank of  $\{\mathbf{\tilde
    H}{\mathbf{Q}}\}_{(N_E \times r_0)} \mathbf{P}^{-1}$ is the same as rank of 
    $\mathbf{\tilde H}_{(N_E \times r_0)}$. Therefore we can simply replace $\{\mathbf{\tilde
    H}{\mathbf{Q}}\}_{(N_E \times r_0)} \mathbf{P}^{-1}$ with ${\mathbf{\tilde
      H}}_{(N_E \times r_0)}$. Thus it suffices to consider the following channel model instead:
\begin{myalign}
  & \mathbf{Y}_t(i) = \mathbf{\Sigma}_{t}  \mathbf{X}_{(r_0 \times 1)} (i)  + \mathbf{Z}_t(i), t=1,2 \label{eq:sig-decomp} \\
  & \mathbf{\tilde Y}(i) = {\mathbf{\tilde H}_{(N_E \times r_0)} \mathbf{X}_{(r_0 \times 1)}(i) } 
\end{myalign}
subject to the following constraint:
\begin{myalign}
E\left[ \frac{1}{ n}\mathrm{trace}\left\{
    {\left( {\mathbf{P} \mathbf{X}^{ n} } \right)\left( { \mathbf{P}
          \mathbf{X}^{ n} } \right)^H } \right\}\right] \le \bar P \label{eq:powerconstraint2}
\end{myalign}
where  $\mathbf{P} \mathbf{X}^{ n}$ denotes the vector formed by concatenating $\{\mathbf{P} \mathbf{X}(i)\}_{1\le i \le n}$.
Recall that if $s_{\mathbf{P}}^2$ denotes the largest eigenvalue of $\mathbf{P}^H
\mathbf{P}$ then we have that:
\begin{myalign}
  \mathrm{trace}\{ (\mathbf{P} \mathbf{X}(i)) (\mathbf{P}
  \mathbf{X}(i))^H \} \le s_{\mathbf{P}}^2
  \mathrm{trace}\{\mathbf{X}(i)\mathbf{X}(i)^H\}.
\end{myalign}
Hence when designing achievable scheme, we use the following power
constraint, which is a sufficient condition for
\eqref{eq:powerconstraint2} to hold:
\begin{myalign}
E\left[ \frac{1}{ n}  \mathrm{trace}(\mathbf{X}^{ n}
  (\mathbf{X}^{ n})^H) \right]\le \frac{\bar P}{s_{\mathbf{P}}^2} \label{eq:powerconstraint3}  
\end{myalign}

We further reduce the channel~\eqref{eq:sig-decomp} to obtain equivalent parallel channels. 
Let $s_{\min}$ be the minimal nonzero element among all diagonal
elements in $\mathbf{S}_1$ and $\mathbf{S}_2$. Replace all diagonal
nonzero elements of $\mathbf{\Sigma}_t$ with $s_{\min}$ and let the
resulting matrix be $\mathbf{\bar \Sigma}_t $. 
We present our coding  scheme for the following channel:
\begin{myalign}
  & \mathbf{Y}_t(i) = \mathbf{\bar \Sigma}_{t}  \mathbf{X}_{(r_0 \times 1)} (i)  + \mathbf{Z}_t(i), \quad t=1,2  \label{eq:model1a}\\
  & \mathbf{\tilde Y}(i) = {\mathbf{\tilde H}_{(N_E \times r_0)}(i)
    \mathbf{X}_{(r_0 \times 1)}(i) } \label{eq:model1b}
\end{myalign}
If we let:
\begin{myalign}
\mathbf{X}_{(r_0  \times 1)} (i) = \left[ \begin{array}{l}
 \mathbf{X}_{\tilde r_1} (i) \\ 
 \mathbf{X}_{s} (i) \\ 
 \mathbf{X}_{\tilde r_2} (i) \\ 
 \end{array} \right]\begin{array}{*{20}c}
   {\;\} \tilde r_1 \;{\rm{rows}}}  \\
   {{\rm{\} }}s\;{\rm{rows}}}  \\
   {\;{\rm{\} }}\tilde r_2 \;{\rm{rows}}}  \\
\end{array}
   \label{eq:x-part}
\end{myalign}
then the two equations given in~\eqref{eq:model1a} reduce to the following
\begin{myalign}
{\mathbf{Y}}_1(i) = s_{\min} \begin{bmatrix} {\mathbf X}_{\tilde r_1}(i) \\ {\mathbf X}_s(i) \end{bmatrix} + {\mathbf Z}_1(i),\label{eq:par-chan-1}\\
{\mathbf{Y}}_2(i) = s_{\min} \begin{bmatrix} {\mathbf X}_s(i) \\ {\mathbf X}_{\tilde r_2}(i) \end{bmatrix} + {\mathbf Z}_2(i).\label{eq:par-chan-2}
\end{myalign}
Thus~\eqref{eq:par-chan-1} and~\eqref{eq:par-chan-2} denote a collection of parallel, independent and identically  distributed channels between the transmitter and the legitimate
receivers. The channels with input  ${\mathbf X}_s(i)$ denote the common channels observed by both  receivers, whereas the channels with input ${\mathbf X}_{\tilde r_1}(i)$ and ${\mathbf X}_{\tilde r_2}(i)$ are only observed by receivers $1$ and $2$ respectively. 

\section{Proof of Theorem~\ref{thm:mimobcsdof}}
\label{sec:thm1}
We first present the key ideas in the coding scheme for a special example, and then present the coding 
scheme for the general case.

\subsection{A Motivating Example: $3\times 2 \times 2 \times 1$ Channel}
\label{sec:structuredbinning}
Consider the special case when $N_T = 3$,
$N_E=1$ and $N_{R_1} = N_{R_2} =2$. For simplicity, we assume that the common message $W_0$ is not present. 
Assume that $r_1 = r_2=2$ and
$r_0=3$ and all the channel matrices are full rank.  
Following the reduction in~\eqref{eq:par-chan-1} and~\eqref{eq:par-chan-2}, 
the channel matrices of the two legitimate receivers reduce to:
\begin{myalign}
& \mathbf{H}_1=[\mathbf{I}_{(2 \times 2)},\mathbf{0}_{(2 \times 1)}],\quad \mathbf{H}_2=[\mathbf{0}_{(2 \times 1)}, \mathbf{I}_{(2 \times 2)}]
\end{myalign}
while the effective channel matrix of the eavesdropper is an arbitrary
rank one matrix.  Assume that we do not impose
a mutual secrecy constraint and let $d_0=0$. Thus according to Theorem~\ref{thm:mimobcsdof}
we seek to achieve $d_1 = d_2 = 1$.

Recall that in \eqref{eq:mainchannel}, the vector $\mathbf{X}$ denotes the transmitter
input. Since the transmitter has three
antennas, i.e., $N_T=3$, $\mathbf{X}$ has three components. To achieve $d_1=1$,
a single-user wiretap codebook ${\cal{C}}_1$ 
for user $1$  requires transmission over the first and the second component of $\mathbf{X}$.
Likewise to achieve $d_2=1$, a single-user wiretap codebook ${\cal{C}}_2$ requires
transmission over the second and third component. Thus the two codebooks must share 
the second component of $\mathbf{X}$. 
However, since $W_1$ and $W_2$ are independent, the signals
that ${\cal{C}}_1$ uses to represent $W_1$ over the second component
of $\mathbf{X}$ in general do not agree with the signals that
${\cal{C}}_2$ uses to represent $W_2$ over this component, causing a
conflict. Thus we cannot simultaneously achieve
$d_1=1$ and $d_2=1$ using this approach.

Our proposed scheme in Theorem~\ref{thm:mimobcsdof} resolves this
conflict by constructing three codebooks, one for each component of
$\mathbf{X}$.  A codebook for the second component of $\mathbf{X}$,
$\cC_E$, is used to transmit a fictitious message $W_E$ via a codeword
$X_E^n(W_E)$.  An independent codebook  $\cC_1$ 
is used to jointly encode $(W_E, W_1)$ into a
codeword $X_1^n(W_E, W_1)$ which is transmitted over the first component of
$\mathbf{X}$. Another codebook $\cC_2$ for the
third component of $\mathbf{X}$ is used to transmit a codeword
$X_2^n(W_E, W_2)$.   Through  random coding analysis, as will be discussed later,
one can show that users $1$ and $2$ can decode $(W_1, W_E)$ and $(W_2, W_E)$
upon observing $(X_1^n, X_E^n)$ and $(X_2^n, X_E^n)$ respectively, with high 
probability. Furthermore as will be shown in the sequel, the messages ($W_1$,
$W_2$) remain simultaneously confidential from any eavesdropper with a single receive antenna. 


To summarize the above example, note that the naive extension of the single-user codebook involves {\em independent}
randomization in the codebooks of the two users. This effectively injects  fictitious 
messages of a higher rate and in turn reduces the message rate. In contrast the proposed
scheme introduces a fictitious message of minimum possible rate needed to guarantee secrecy.

In generalizing the above example to arbitrary number of antennas, we use three i.i.d.\ Gaussian codebooks, and assign a subset of parallel channels for each codeword. The rate of the codebooks is selected such that the average error probability at the legitimate receivers under maximum likelihood decoding is arbitrarily small. We also show in section~\ref{subsec:side-info} that if the information messages are revealed to the eavesdropper, the error probability in decoding the fictitious message given the eavesdropper's observation, also vanishes to zero. For a codebook satisfying these properties, we provide the secrecy analysis in section~\ref{sec:secrecy} and complete the proof of the coding theorem. We note that this approach of secrecy analysis, where the eavesdropper is able to decode the fictitious message given side information, is routinely used when establishing the achievability of weak-secrecy.

\subsection{Achievability}
\label{sec:achievablity}
Since the secrecy rate is  zero whenever $r_0 \le N_E$
(c.f.~\cite{xiang2010arb}), without loss of generality, we assume $r_0> N_E$
and consider $\mathbf{\tilde H}$ that has the following form:
\begin{myalign}
  \mathbf{\tilde H}=[\mathbf{I}_{N_E \times
    N_E},\mathbf{0}_{N_E \times (r_0-N_E)}]\mathbf{U}_E \stackrel{\Delta}{=} {\mathbf{\tilde{U}}}_E\label{eq:normalize}
\end{myalign}
where $\mathbf{U}_E$ is a unitary matrix, which is only known by the
eavesdropper, and ${\mathbf{\tilde{U}}}_E$ represents the first $N_E$ rows of $\mathbf{U}_E$. 
Furthermore recall that the legitimate receiver's channel are parallel independent broadcast channels:
\begin{myalign}
{\mathbf{Y}}_1(i) = s_{\min} \begin{bmatrix} {\mathbf X}_{\tilde r_1}(i) \\ {\mathbf X}_s(i) \end{bmatrix} + {\mathbf Z}_1(i)\label{eq:par-chan-11}\\
{\mathbf{Y}}_2(i) = s_{\min} \begin{bmatrix} {\mathbf X}_s(i) \\ {\mathbf X}_{\tilde r_2}(i) \end{bmatrix} + {\mathbf Z}_2(i)\label{eq:par-chan-21}
\end{myalign}

As in \cite{goel2008guaranteeing, xiang2010arb}, we then introduce
artificial noise into $\mathbf{X}$ as:
\begin{myalign}
  \mathbf{X}(i)=\mathbf{\bar X}_{(r_0 \times 1)}(i)+\mathbf{N}(i) \label{eq:def_tilde_X}
\end{myalign}
where $\mathbf{N}$ is the $r_0 \times 1$ artificial noise vector
consisting of independent rotationally invariant complex Gaussian
random variables with zero mean and unit variance. In contrast $\mathbf{\bar X}$
is the information bearing signal which will be used in the codebook transmission.

 Let $P$ be
such that $P=\frac{\bar P}{s_{\mathbf P}^2}-r_0$ (c.f.~\eqref{eq:powerconstraint3}). We shall allocate a total power of $r_0$ units on artificial noise
$\mathbf{N}$ in \eqref{eq:def_tilde_X} and $P$ units on $\mathbf{\bar X}$. Let the rate $R$  be defined by:
\begin{align}
R = C({s^2_{\min}(P/r_0)}/({s^2_{\mathrm{min}}+1})) \label{eq:R0-def}
\end{align} 
where $C\left( x \right) \stackrel{\Delta}{=} \log _2 \left( {1 + x} \right)$.
Note that $R$ is the rate supported over each parallel channel in~\eqref{eq:par-chan-11} and~\eqref{eq:par-chan-21}.

We sample our codebooks from an i.i.d.\ Gaussian random ensemble as discussed next.
Let $\bar \epsilon > 0$ be a fixed constant. Let $Q_{k}(\xvec)$ denote the $k$-dimensional rotationally invariant
complex Gaussian distribution with covariance matrix $(P(1- \bar
\epsilon)/r_0) \mathbf{I}_{(k \times k)}$ where $\mathbf{I}_{k\times k}$  denotes an identity matrix of dimension $k$. Define the $n$-letter Gaussian input distribution $Q_{k}(\xvec^n)$ as
$ Q_{k}(\xvec^n) =  \prod\limits_{i = 1}^n {Q_{k}\left( {\xvec_i } \right)}$.
In the following we consider transmission of four messages $W_i \in [1, 2^{n d_i R}], $ $W_0 \in [1, 2^{n d_0 R}]$
 and $W_E \in [1, 2^{n N_E R}]$, where $W_0$ is the common message and  $W_1$ and $W_2$ are the private messages
 that need to be decoded by receivers 1 and 2 respectively. The message $W_E$ is a fictitious message. 
As in~\eqref{eq:inthiscase} we define $N_E'$ using:
\begin{myalign}
  N_E'=N_E+d_0. \label{eq:NE'-def}
\end{myalign}
Such a notation is again convenient, since in our coding scheme we will jointly code the message pair $(W_E, W_0)$ of a total rate $2^{nR(N_E+d_0)}$. 
We separately consider the case when $s \le N_E'$ and when $s > N_E'$.
\begin{figure}[t]
  \centering 
\includegraphics[width=\columnwidth]{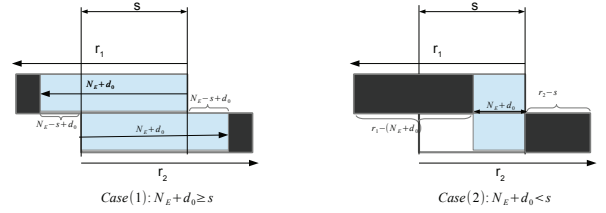}
  \caption{Codebook generation:  (1) $s \le N_E + d_0 \le \min\{r_1, r_2\}$
    (2) $0\le N_E+d_0 <s$.  Here $s=r_1+r_2-r_0$ denotes the dimension of the common subspace. The shaded blue region indicates the dimensions where the common and fictitious messages i.e., $\bar{\bf X}_B^n(W_0,W_E)$ are transmitted. The shaded black portion indicate the dimensions where the private messages are transmitted.   In case (1), in addition to the common subspace, we further need to use $(N_E+d_0 -s)$ dimensions of each private subspace for transmitting the common message. In case (2), we have a surplus $(s-N_E-d_0)$ dimensions in the common subspace. These can be used for transmitting private messages. Therefore the sum-rate constraint in Theorem~\ref{thm:mimobcsdof} is active in this case.}
  \label{fig:codebook}
\end{figure}

 
\subsubsection{Case 1  ($s \le N_E' \le \min(r_1, r_2)$)}
\label{s-small}

In this case recall from  Theorem~\ref{thm:mimobcsdof} and Fig.~\ref{fig:dim} that the $(d_1, d_2)$ region is a rectangle. It suffices to show that
any triple $(\td_0, \td_1, \td_2)$  such that  $\td_0 \le N_E' - N_E$  and $\td_i \le r_i - N_E'$ for $i=1,2$, is achievable.
Following the decomposition  illustrated in Figure~\ref{fig:codebook} let:
\begin{myalign}
\mathbf{\bar X}_{(r_0  \times 1)} (i) = \left[ \begin{array}{l}
 \mathbf{\bar X}_A (i) \\ 
 \mathbf{\bar X}_B (i) \\ 
 \mathbf{\bar X}_C (i) \\ 
 \end{array} \right]\begin{array}{*{20}l}
   {\}  r_1-N_E'\;{\rm{ rows}}}  \\
   {{\rm{\} }}2N_E'-s\;{\rm{rows}}}  \\
   {{\rm{\} }}r_2-N_E'\;{\rm{ rows}}}  \\
\end{array}  \label{eq:x-case_one}
\end{myalign}
We let the three components above correspond to three different codebooks $\cC_A,$ $\cC_B$ and $\cC_C$ indicated below.
\begin{itemize}
\item The codebook ${\mathcal C}_B$ maps the message-pair $(W_0,W_E)$
  to a codeword ${\mathbf{\bar{X}}}_B^n(W_0,W_E)$. It consists of
  $2^{n (N_E' R)}$ codewords. Each codeword is sampled in an
  i.i.d. fashion from the distribution $Q_{(2N_E'-s) }(\xvec)$. 
   The codeword is transmitted through the component $\mathbf{\bar
    X}_B$ in~\eqref{eq:x-case_one} as discussed below.
\begin{align}
\mathbf{\bar X}_B(i) =\left[ \begin{array}{l}
 \mathbf{\bar X}_{B1} (i) \\ 
 \mathbf{\bar X}_{B0} (i) \\ 
 \mathbf{\bar X}_{B2}(i) \\ 
 \end{array} \right]\begin{array}{*{20}l}
   {\}  N_E'-s\;{\rm{ rows}}}  \\
   {{\rm{\} }}s\;{\rm{rows}}}  \\
   {{\rm{\} }}N_E'-s\;{\rm{ rows}}}  \\
\end{array}
\end{align}
and let $\mathbf{\bar X}_s(i)=  \mathbf{\bar X}_{B0} (i),$  and furthermore
\begin{align}
\mathbf{\bar X}_{\tilde{r}_1}(i) =\left[ \begin{array}{l}
 \mathbf{\bar X}_{A} (i) \\ 
 \mathbf{\bar X}_{B1} (i) \\ 
 \end{array} \right] \quad
 \mathbf{\bar X}_{\tilde{r}_2}(i) =\left[ \begin{array}{l}
 \mathbf{\bar X}_{B2} (i) \\ 
  \mathbf{\bar X}_{C} (i) \\ 
 \end{array} \right],
\end{align}
where the vectors $\mathbf{\bar X}_s(i),$ $\mathbf{\bar X}_{\tilde{r}_1}(i)$ and $\mathbf{\bar X}_{\tilde{r}_2}(i)$
are  inputs into the parallel channels in~\eqref{eq:x-part}. 

\item The codebook ${\mathcal C}_A$ maps a message pair $(W_0, W_1,
  W_E)$ to a codeword ${\mathbf{\bar{X}}}_A^n (W_0, W_1, W_E)$. It
  consists of a total of $2^{n( (\td_0 + \td_1 + N_E) R)}$ codewords each
  sampled in an i.i.d. fashion from the distribution $Q_{(r_1-N_E') }(\xvec)$. 
  The codeword will be transmitted through $\mathbf{\bar
    X}_A$.

\item The codebook ${\mathcal C}_C$ maps the message pair $(W_0, W_2,
  W_E)$ to a codeword ${\mathbf {\bar{X}}}_C^n (W_0, W_2, W_E)$. It
  consists of a total of $2^{n( (\td_0 + \td_2 + N_E) R)}$ codewords. Each codeword is
  sampled in an i.i.d. fashion from the distribution $Q_{
  (r_2-N_E')}(\xvec)$. 
\end{itemize}

Given a message pair $(W_0, W_1, W_2, W_E)$ the encoder generates the
associated sequence $\bar{\mathbf X}^n$ (c.f.~\eqref{eq:x-case_one})
and transmits ${\mathbf X}^n$ (c.f.~\eqref{eq:def_tilde_X}) over $n$
channel uses. We declare an error if ${\mathbf X}^n$  does not satisfy the average power constraint
(cf.~\eqref{eq:powerbc}). By selecting $n$ to be sufficiently large, this error can be made arbitrarily small. 

The received signal ${\mathbf Y}_1^n$~(c.f.~\eqref{eq:par-chan-1}) at receiver $1$ can be expressed as:
\begin{myalign}
\mathbf{Y}_1 (i) = 
\left[ \begin{array}{l}
 \mathbf{Y}_{A} (i) \\ 
 \mathbf{Y}_{B1} (i) \\ 
  \mathbf{Y}_{s} (i) \\ 
 \end{array} \right]\begin{array}{*{20}l}
   {\} r_1-N_E '\mathrm{\;rows}}  \\
   {\} N_E ' - s \mathrm{\;rows}}  \\
   {\} s \mathrm{\;rows}}  \\
\end{array}.\label{eq:rewriteY1}
\end{myalign}
Receiver $1$ decodes $(W_0,W_1,W_E)$ in the following order:
\begin{enumerate}
\item Decode $(W_0,W_E)$ from $(\mathbf{Y}_{B1}^n, \mathbf{Y}_{s}^n)$ using a maximal likelihood decoder:
  \begin{myalign}
    (\hat W_0,\hat W_E)=\arg \mathop {\max }\limits_{w_0 ,w_E } \Pr \left( {\mathbf{\bar{X}}}_{B}^n (w_0, w_E)|\mathbf{Y}_{B1}^n, \mathbf{Y}_{s}^n\right)\label{eq:MLDecoder1}
  \end{myalign}
\item Decode $W_1$ from $\mathbf{Y}_A^n$ using a maximal likelihood decoder:
  \begin{myalign}
    \hat W_1=\arg \mathop {\max }\limits_{w_1 } \Pr \left( {\mathbf{\bar{X}}}_{A}^n (\hat W_0, w_1, \hat W_E)|\mathbf{Y}_{A}^n \right)\label{eq:MLDecoder2}
  \end{myalign}
\end{enumerate}
It can be shown through standard analysis\footnote{ If a joint-typicality based decoder is used we still obtain the same rate. However the maximum likelihood decoder also guarantees that the error probability approaches zero exponentially with the block-length\cite[(7.3.22)]{gallager1968information}. This particular scaling is useful in showing the existence of a single universal codebook that remains confidential against all eavesdropper channels simultaneously as done in the single user case~\cite{xiang2010arb}. } that the error probability in~\eqref{eq:MLDecoder1} approaches zero provided $(R_0,R_E)$
satisfy the following:
\begin{align}
R_0 + R_E &< I({\mathbf{\bar{X}}}_B; {\mathbf Y}_{B1},{\mathbf Y}_{s}) = N_E' R
\end{align}
where the rate $R$ is the rate associated with each parallel channel~\eqref{eq:R0-def}. This shows that any $\td_0 \le  N_E' - N_E =d_0$ (c.f.~\eqref{eq:NE'-def})
is achievable at user $1$. Furthermore the error probability in~\eqref{eq:MLDecoder2} vanishes to zero provided that
\begin{align}
R_1 &<  I({\mathbf{\bar{X}}}_A; {\mathbf Y}_{A}) = (r_1 - N_E')R
\end{align}
is satisfied i.e.,  ${\td_1 \le r_1-N_E'}$ is achievable for user $1$. In an analogous manner we can show that $\td_0 = N_E'-N_E$
and $\td_2 = r_2-N_E'$ are achievable for user $2$.

\subsubsection{Case 2 ($N_E' < s$)}
\label{s-large}
In this case the sum-rate constraint in Theorem~\ref{thm:mimobcsdof} is active.
We show that $\td_0 \le N_E' - N_E$ as well as the corner point $(\td_1, \td_2)=(r_1-N_E', r_2-s)$ is
achievable. By a symmetric argument it follows that the corner point
$(r_1-s, r_2-N_E')$ is also achievable. The achievability of the
entire region then follows using a time-sharing argument.

To define our code construction, we begin by
splitting the input symbols ${\mathbf{ \bar X}}_s$ in~\eqref{eq:x-part}  into two groups:
\begin{myalign}
\mathbf{\bar X}_s  = \left[ {\begin{array}{*{20}c}
   {\mathbf{\bar X}_{s1} }  \\
   {\mathbf{\bar X}_{s2} }  \\
\end{array}} \right]\begin{array}{*{20}l}
   {\} s - N'_E \;{\rm{rows}}}  \\
   {\} N'_E \;{\rm{rows}}}  \\
\end{array}
\end{myalign}
Define $\mathbf{\bar X}_{A}$ as
\begin{myalign}
\mathbf{\bar X}_{A}  = \left[ \begin{array}{l}
 \mathbf{\bar X}_{\tilde r_1}  \\ 
 \mathbf{\bar X}_{s1}  \\ 
 \end{array} \right]\begin{array}{*{20}l}
   {\} \tilde r_1\;{\rm{rows}}}  \\
   {{\rm{\} }}s - N'_E\;{\rm{ rows}}}  \\
\end{array},
\end{myalign}where $ \mathbf{\bar X}_{\tilde r_1}$ constitutes the input to parallel channels of receiver $1$ in
\eqref{eq:x-part}.  Let $\mathbf{\bar X}_B=\mathbf{\bar
  X}_{s2}$ and $\mathbf{\bar X}_C=\mathbf{\bar X}_{\tilde
  r_2}$. The overall input vector is expressed via ${\mathbf{\bar X}} = \left[\begin{array}{c} {\mathbf{\bar X}}_A \\
  {\mathbf{\bar X}}_B \\ {\mathbf{\bar X}}_C \end{array}\right].$
  
As before we use $\mathbf{\bar X}_{A}$ to send $W_1$ to user $1$, 
use $\mathbf{\bar X}_C$ to send $W_2$ to user $2$, and use the second group ${\mathbf{\bar X}}_{B}$ for sending $W_0$ and $W_E$. The associated codebook construction is discussed next.
\begin{itemize}
\item The codebook ${\mathcal C}_B$ that maps the message pair $(W_0,
  W_E)$ to a codeword ${\mathbf{\bar{X}}}_{B}^n(W_0, W_E)$.  It
  consists of $2^{n (N_E' R)}$ codewords. Each codeword is 
   sampled in an i.i.d. fashion from the
  distribution $Q_{N_E}(\xvec)$.

\item The codebook ${\mathcal C}_{A}$ that maps each message pair
  $(W_0, W_1, W_E)$ to a codeword 
  
  ${\mathbf{\bar{X}}}_{A}^n(W_0,
  W_1, W_E)$. It consists of a total of $2^{n( (r_1-N_E') R)}$
  codewords. Each codeword is 
   sampled in an i.i.d. fashion from the distribution
  $Q_{(r_1-N_E') }(\xvec)$.

\item The codebook ${\mathcal C}_C$ maps the message pair $(W_0, W_2,
  W_E)$ to a codeword ${\mathbf{\bar{X}}}_{C}^n(W_0, W_2, W_E)$. It
  consists of a total of $2^{n( (r_2-s) R)}$ codewords each sampled in
  an i.i.d. fashion from the distribution $Q_{(r_2-s)}(\xvec)$.

\end{itemize}
Upon receiving ${\mathbf Y}_1^n$~(c.f.~\eqref{eq:par-chan-1}),
receiver $1$ decomposes each of its component $\mathbf{Y}_1(i)$ as:
\begin{myalign}
\mathbf{Y}_1 (i) = 
\left[ \begin{array}{l}
 \mathbf{Y}_{A} (i) \\ 
  \mathbf{Y}_{s} (i) \\ 
 \end{array} \right]\begin{array}{*{20}l}
   {\} r_1-N_E '\mathrm{\;rows}}  \\
   {\} N_E' \mathrm{\;rows}}  \\
\end{array}.\label{eq:rewriteY2}
\end{myalign}

Receiver $1$ decodes $(W_0,W_1,W_E)$ in the following order:
\begin{enumerate}
\item Decode $(W_0,W_E)$ from $\mathbf{Y}_{s}^n$ using the maximal likelihood decoder:
  \begin{myalign}
    (\hat W_0,\hat W_E)=\arg \mathop {\max }\limits_{w_0 ,w_E } \Pr \left( {\mathbf{\bar{X}}}_{B}^n (w_0, w_E)|\mathbf{Y}_{s}^n\right)\label{eq:MLDecoder12}
  \end{myalign}

\item Decode $W_1$ from $\mathbf{Y}_A^n$ using the maximal likelihood decoder in \eqref{eq:MLDecoder2}.
\end{enumerate}
It can be shown that the error probability associated with~\eqref{eq:MLDecoder12} vanishes to zero if 
$R_0 + R_E < I({\mathbf{\bar{X}}}_{B}; \mathbf{Y}_{s}) = N_E' R$. This shows that $\td_0 \le N_E' - N_E =d_0$ is achievable.
Likewise the error probability associated with message $W_1$ vanishes to zero if $R_1 < I({\mathbf{\bar{X}}}_{A}; \mathbf{Y}_{A}) = (r_1-N_E')R$.
This shows that $\td_1 \le r_1- N_E'$ is achievable for user $1$.

In an analogous matter we can show that the error probability at user $2$ vanishes to zero if $\td_0 \le N_E' - N_E$ and $\td_2 \le r_2-s$.


\subsection{ Side Information Assisted Decoding at Eavesdropper}
\label{subsec:side-info}
We next argue that in the codebook ensemble $({\mathcal C}_A \times {\mathcal C}_B \times {\mathcal C}_C )$ there exists at-least one codebook such that the eavesdropper can also reliably decode 
$W_E$ given $(W_0, W_1, W_2)$. Such a codebook will be used in the analysis of equivocation in the next sub-section. Our proposed decoder is also a maximum likelihood decoder as stated below:
  \begin{align}
  \hat{W}_E &= \arg\max_{w_E} \Pr\bigg({\mathbf{\bar{X}}}_A^n(\hat{W}_0, \hat{W}_1, w_E),{\mathbf{\bar{X}}}_B^n(\hat{W}_0, w_E), \notag\\ &\qquad\qquad\qquad{\mathbf{\bar{X}}}_C^n(\hat{W}_0, \hat{W}_2, w_E)  | {\mathbf{\tilde{Y}}_{{\mathbf{\tilde{h}}}}}^n \bigg)
  \end{align}
  where $(\hat{W}_0, \hat{W}_1, \hat{W}_2)$ denote the messages revealed to the eavesdropper. 
The error probability decays to zero  if the rate  $R_E$ satisfies:
\begin{align}
R_E &< I({\mathbf{\bar{X}}}_A,{\mathbf{\bar{X}}}_B,{\mathbf{\bar{X}}}_C  ; {\mathbf{\tilde{Y}}}_{\mathbf{\tilde{h}}})\\
&=I({\mathbf{\bar{X}}}_A,{\mathbf{\bar{X}}}_B,{\mathbf{\bar{X}}}_C  ; {\mathbf{\tilde{U}}}_{E} {\mathbf{\bar{X}}}+ {\mathbf{\tilde{U}}}_{E}{\mathbf N} )\label{eq:uE}\\
&= \log\det\left({\mathbf{I}} + \frac{P}{r_0} {\mathbf{\tilde{U}}}_{E} {\mathbf{\tilde{U}}}_{E}^H\right)\label{eq:log-det}\\
&= N_E \log\left(1+ \frac{P}{r_0}\right)\label{eq:Re-bnd}
\end{align}
where ${\mathbf{\tilde{U}}}_{E}$
denotes the first $N_E$ rows of the unitary matrix ${\mathbf{U}}_E$ (c.f.~\eqref{eq:normalize}).  We further substitute~\eqref{eq:def_tilde_X}  in~\eqref{eq:uE},~\eqref{eq:log-det} follows from the fact that the entries of ${\mathbf{\tilde{X}}}$
are sampled i.i.d.\ from ${\mathcal CN}(0, P/r_0)$ and the last relation uses ${\mathbf{\tilde{U}}}_{E} {\mathbf{\tilde{U}}}_{E}^H={\mathbf I}_{N_E}$.
Since we select $R_E = N_E R$, where $R$  (c.f.~\eqref{eq:R0-def}) satisfies $R \le  \log\left(1+ \frac{P}{r_0}\right)$, it follows~\eqref{eq:Re-bnd} is indeed satisfied. This shows that in the ensemble of codebooks there exists at-least one codebook such that $W_E$ is reliably decoded by the eavesdropper with a fixed channel matrix ${\mathbf{\tilde{h}}}$. 
Thus using Fano's inequality:
\begin{align}
H\left( {W_E |\mathbf{\tilde Y}^n_{\mathbf{\tilde h}} ,W_0,
      W_1 ,W_2 } \right) \le n\epsilon_n  \label{eq:ent-bnd-0}
\end{align}
for some sequence $\epsilon_n$ that converges to zero as $n\rightarrow\infty$.
Using a standard union bound argument it follows that there is at-least one codebook which can be reliably decoded by the legitimate receivers and that satisfies~\eqref{eq:ent-bnd-0}.

In order to establish the secrecy constraint~\eqref{eq:secrecybc}, we need to demonstrate  that there exists a single codebook that satisfies~\eqref{eq:ent-bnd-0} for every possible realization of the eavesdropper channel matrix $\mathbf{\tilde H},$ whose rank equals $N_E$. The existence of such codebooks generally follows from the 
 compound channel coding theorem~\cite[Theorem 7.1 (pp.~170), Remark 7.3  (pp.~172)]{elGamal}, \cite[Eq.~(11)]{koksal}.
We remark that when the set of possible states  $\mathbf{\tilde H}$ is finite, the proof of the compound channel coding theorem exploits a  union bound argument over the states. In the present case $\mathbf{\tilde H}$ belongs to a continuous set. Therefore a simple union bound argument cannot be applied. Suitable quantization of the channel matrices is needed to show that the error probability simultaneously goes to zero for each state. We refer the reader to~\cite{xiang2010arb} where such an argument was carefully outlined  for the single-user case, but leave out  a detailed argument in the present paper as it is analogous\footnote{One key difference required in the extension to continuous set of channels is that we cannot sample the codewords i.i.d.\  and then use expurgation to satisfy the power constraint as is commonly done (see e.g.,~\cite[pp.~243-245]{cover}). Instead we need to sample codewords from a normalized distribution, so that they lie within a ball. We refer the reader to~\cite{xiang2010arb} for further details. }. We note that  such a codebook guarantees that~\eqref{eq:ent-bnd-0} is satisfied for any $\mathbf{\tilde H}$ whose rank equals $N_E$ i.e., 
\begin{align}
\sup_{{\mathbf{\tilde h}}:~\text{rank}({\mathbf{\tilde h}}) = N_E}H\left( {W_E |\mathbf{\tilde Y}^n_{\mathbf{\tilde h}} ,W_0,
      W_1 ,W_2 } \right) \le n\epsilon_n.  \label{eq:ent-bnd-00}
\end{align}

\subsection{Equivocation analysis}
\label{sec:secrecy}

It suffices to demonstrate the secrecy constraint when the rank of  $\mathbf{\tilde H}$ equals $N_E$.  If the
rank is smaller than $N_E$ the secrecy constraint clearly holds for
this weaker eavesdropper. We shall only present the secrecy analysis for $N_E' \ge s$. The analysis for $N_E' <s$ is completely analogous.
As discussed in section~\ref{subsec:side-info}, we consider a codebook that  satisfies~\eqref{eq:ent-bnd-00}. 
\begin{myalign}
& H\left( {W_0, W_1 ,W_2 |\mathbf{\tilde Y}_{\mathbf{\tilde h}}^n } \right)  \label{eq:beforeapplyFano}
\ge I\left( {W_0, W_1 ,W_2 ;\mathbf{\bar X}^n |\mathbf{\tilde Y}_{\mathbf{\tilde h}}^n } \right) \\ 
  =& H\left( {\mathbf{\bar X}^n |\mathbf{\tilde Y}_{\mathbf{\tilde h}}^n }
  \right) - H\left( {\mathbf{\bar X}^n |\mathbf{\tilde Y}^n_{\mathbf{\tilde h}} ,W_0,
      W_1 ,W_2 } \right) \label{eq:applyFano}\\
        =& H\left( {\mathbf{\bar X}^n |\mathbf{\tilde Y}_{\mathbf{\tilde h}}^n }
  \right) - H\left( {W_E |\mathbf{\tilde Y}^n_{\mathbf{\tilde h}} ,W_0,
      W_1 ,W_2 } \right) \label{eq:applyFano-1} \\
        =& H\left( {\mathbf{\bar X}^n} \right) - I\left({\mathbf{\bar X}^n}; {\mathbf{\tilde Y}_{\mathbf{\tilde h}}^n }
  \right) - H\left( {W_E |\mathbf{\tilde Y}^n_{\mathbf{\tilde h}} ,W_0, 
      W_1 ,W_2 } \right) \label{eq:applyFano-2}
\end{myalign}
where~\eqref{eq:applyFano-1} follows from the fact that ${\bar{\mathbf X}^n}$ is a deterministic function of $(W_0, W_E, W_1, W_2)$.

On the other hand since rank$(\mathbf{\tilde h}) = N_E$ it can be shown that:
\begin{myalign}
w\left(\mathop {\lim }\limits_{n\to \infty }  {\frac{1}{n} I\left(
    {\mathbf{\bar X}^n ;\mathbf{\tilde Y}_{\mathbf{\tilde h}}^n } \right)}\right) \le
N_E.
\label{eq:ent-bnd-1}
\end{myalign}
For the first term in \eqref{eq:applyFano-2}, since $H( {\mathbf{\bar X}^n }) = H(W_0, W_1, W_2, W_E)$ we have:
\begin{myalign}
w\left(\lim_{n\rightarrow\infty} \frac{1}{n}H( {\mathbf{\bar X}^n }) \right)= \td_0 + \td_1 + \td_2 + N_E 
\label{eq:ent-bnd-2}
\end{myalign}
Applying~\eqref{eq:ent-bnd-00},~\eqref{eq:ent-bnd-1} and~\eqref{eq:ent-bnd-2} to \eqref{eq:beforeapplyFano}-\eqref{eq:applyFano-2}, we have that
\begin{align}
w\left(\lim_{n\rightarrow \infty} \frac{1}{n}\sup_{\mathbf{\tilde h}}H(W_0, W_1, W_2 | {\mathbf Y}_{\mathbf{\tilde h}}^n)\right) \ge \td_0 + \td_1 + \td_2
\label{eq:ent-bnd-3}
\end{align}
as required. This completes the proof of our coding theorem.
\subsection{Converse}
\label{sec:converse}
We now establish the  upper bounds stated in~\eqref{eq:bnd-t1} and~\eqref{eq:sumratebound}. The two upper bounds in~\eqref{eq:bnd-t1} are single-user bounds whereas the upper bound in~\eqref{eq:sumratebound} involves the sum-rate. These bounds correspond to the three cuts in the broadcast network~\cite{elGamal}. For each cut, as discussed below we find the worst case eavesdropper channel. 

Recall that the rank of $\mathbf{H}_1$ is $r_1$. To
establish~\eqref{eq:bnd-t1}, we express (after row permutation if necessary) $\mathbf{H}_1$ as
\begin{align}
\mathbf{H}_1  = \left[ \begin{array}{l}
 \mathbf{H}_{11}  \\ 
 \mathbf{H}_{12}  \\ 
 \end{array} \right]\begin{array}{*{20}l}
   {\} N_E \;{\rm{rows}}}  \\
   {\} N_{R_1 }  - N_E\;{\rm{ rows}}}  \\
\end{array}.
\end{align}
such that the matrix $\mathbf{H}_{12}$ has a rank of $\{r_1-N_E\}^+$. Since the
eavesdropper channel state is arbitrary, we consider an eavesdropper
channel for which $\mathbf{\tilde H} = {\mathbf H}_{11}$ whose rank clearly 
equals $N_E$.  For any
coding scheme we can upper bound the rate as follows:
\begin{myalign}
&n(R_0+R_1) = H(W_0, W_1)\\
&\le I(W_0, W_1; {\mathbf Y}_1^n) + n\epsilon_n \label{eq:Fano-Ineq-1}\\
&\le I(W_0, W_1; {\mathbf Y}_1^n)  -I(W_0, W_1; \mathbf{\tilde Y}^n) + 2n\epsilon_n \label{eq:Equiv}\\
&\le I(W_0, W_1; {\mathbf Y}_1^n, \mathbf{\tilde Y}^n)  -I(W_0, W_1; \mathbf{\tilde Y}^n) + 2n\epsilon_n \label{eq:Equiv-1}\\
&= I(W_0, W_1; {\mathbf Y}_1^n | \mathbf{\tilde Y}^n) + 2n\epsilon_n\\
&\le h({\mathbf Y}_1^n | \mathbf{\tilde Y}^n) - h({\mathbf Y}_1^n | \mathbf{\tilde Y}^n,W_0, W_1) + 2n\epsilon_n\\
&\le h({\mathbf Y}_1^n | \mathbf{\tilde Y}^n) - h({\mathbf Y}_1^n | \mathbf{\tilde Y}^n,W_0, W_1, {\mathbf X}^n) + 2n\epsilon_n\\
&=  I( {W_0, W_1, {\mathbf X}^n; \mathbf Y}_1^n | \mathbf{\tilde Y}^n) + 2n\epsilon_n\\
&=  I(  {\mathbf X}^n; \mathbf{H}_1 {\mathbf X}^n + {\mathbf{Z}}_1^n | \mathbf{H}_{11} \mathbf{X}^n ) + 2n\epsilon_n\label{eq:X-Markov}\\
&\le I({\mathbf X}^n; {\mathbf{H}}_{12}{\mathbf X}^n + {\mathbf{Z}}_{12}^n)+ 2n\epsilon_n \label{eq:mimo-cap}
\end{myalign}
where ${\mathbf{Z}}_{12}$ is the last $N_{R_1}-N_E$ rows of
$\mathbf{Z}_1$. The step \eqref{eq:Fano-Ineq-1} follows from Fano's
inequality since $(W_0, W_1)$ must be decodable by receiver $1$
while~\eqref{eq:Equiv} is a consequence of the equivocation constraint.
Finally~\eqref{eq:X-Markov} follows from the Markov Condition that $(W_0,W_1) \rightarrow {\mathbf X}^n \rightarrow
{\mathbf Y}_1^n$.
Since the right hand side of~\eqref{eq:mimo-cap} is upper bounded by the
capacity of a MIMO Gaussian channel~\cite{tseViswanath} and by our construction, the rank of
${\mathbf{H}}_{12}$ equals $(r_1-N_E),$ it follows that
\begin{align}
d_0 + d_1 \le \left\{r_1-N_E\right\}^+.
\end{align}
In a similar fashion, we can show that
\begin{align}
d_0 + d_2 \le \left\{r_2-N_E\right\}^+.
\end{align}
To establish sum-rate bound on $(W_0, W_1, W_2)$  we let the two users cooperate
and obtain with ${\mathbf H}_{\cup} = \left[\begin{array}{c}{\mathbf{H}}_1 \\ {\mathbf{H}}_2 \end{array} \right]$ and ${\mathbf Z} =  \left[\begin{array}{c}{\mathbf{Z}}_1 \\ {\mathbf{Z}}_2 \end{array} \right]$:
\begin{align}
n(R_0+R_1+R_2) \le I({\mathbf X}^n ;{\mathbf H}_{\cup} {\mathbf X}^n + {\mathbf{Z}}^n | \mathbf{\tilde Y}^n ) + 2n\epsilon_n.
\end{align}
Since the rank of ${\mathbf H}_{\cup}$ equals $r_0 = r_1 + r_2 - s,$ we can express  (after suitable row permutations)
\begin{align}
{\mathbf H}_{\cup} = 
\left[
\begin{array}{c}
{\mathbf A}\\
{\mathbf B}
\end{array}
\right]\begin{array}{*{20}l}
   {\} N_E \;{\rm{rows}}}  \\
   {\} N_{R_1 }  + N_{R_2}- N_E\;{\rm{ rows}}}  \\
\end{array}.\label{eq:Hu}
\end{align}
where the rank of ${\mathbf B}$ equals $(r_0 - N_E)$. We further select $\mathbf{\tilde H} = {\mathbf A}$ of rank $N_E$. 
Using  ${\mathbf{Z}}_A$ and $ {\mathbf{Z}}_B$ to denote the projection of the noise vector $ {\mathbf{Z}}$
onto the rows corresponding to matrices ${\mathbf A}$ and ${\mathbf B}$ respectively, the
 sum-rate can be upper bounded as follows:
\begin{align}
&n(R_0+R_1+R_2) \le I({\mathbf X}^n ;{\mathbf H}_{\cup} {\mathbf X}^n + {\mathbf{Z}}^n | \mathbf{\tilde Y}^n ) + 2n\epsilon_n\\
&= I({\mathbf X}^n ;{\mathbf A} {\mathbf X}^n + {\mathbf{Z}}_A^n,  {\mathbf B} {\mathbf X}^n + {\mathbf{Z}}_B^n | \mathbf{A} {\mathbf X}^n ) + 2n \epsilon_n \label{eq:proj-z}\\
&\le I({\mathbf X}^n ;  {\mathbf B} {\mathbf X}^n + {\mathbf{Z}}_B^n) + 2n\epsilon_n\label{eq:sum-rate-bnd}
\end{align}
Since~\eqref{eq:sum-rate-bnd} is upper bounded by the capacity of a MIMO channel with channel matrix ${\mathbf B}$ of rank $(r_0 - N_E)$
it follows that
\begin{align}
d_0 + d_1 + d_2 \le \left\{r_0 - N_E\right\}^+.
\end{align}
This completes the proof of the converse.

\section{Proof of Theorem~\ref{thm:mimobcsdof-2}}
\label{sec:thm2}
We outline how our results change when an additional mutual privacy constraints across the two receivers are imposed.

For the case when $d_0 + N_E \ge s,$ we  can in fact use the same construction as in Section~\ref{s-small}. Since the message $W_1$ is transmitted only through symbols ${\mathbf{\bar{X}}}_A(i)$ which are not observed by user $2,$ its mutual privacy constraint is clearly satisfied. Similarly message $W_2$ is transmitted only through symbols ${\mathbf{\bar{X}}}_C(i)$ which are not observed by user $1$. Hence its mutual privacy constraint is also satisfied. Clearly the addition of the mutual privacy constraint can only reduce the achievable d.o.f.\ and hence the proposed scheme achieves optimal d.o.f. 

When $d_0 + N_E \le s$ we use construction in section~\ref{s-large} with $N_E' = s$. Thus the codebook $\cC_B$ uses all available ${\mathbf{\bar{X}}}_s$ symbols whereas the codebooks $\cC_A$ and $\cC_C$ use ${\mathbf{\bar{X}}}_{\tilde{r}_1}$ and ${\mathbf{\bar{X}}}_{\tilde{r}_2}$  symbols respectively.  
It can again be readily verified that any pair $(d_0, d_1, d_2)$ that satisfies:  $d_i \le r_i - s$ and $d_0 \le s- N_E$ is achievable.

For the converse we note that the constraint $d_0 + d_i \le r_i - N_E$  follows from section~\ref{sec:converse} which does not use the mutual-privacy constraint. To establish the condition $d_i \le r_i - s,$ we assume that the messages $(W_0, W_2)$ are revealed to the two receivers and only consider the transmission of message $W_1$.  The resulting system reduces to a MIMO wiretap channel where the channel matrices ${\mathbf H}_1$ and ${\mathbf H}_2$ are known to the transmitter. As shown in~\cite{khisti:stm2} the high SNR capacity is achieved using the generalized singular value decomposition and the associated degree of freedom satisfies $d_1 \le r_1-s$. The upper bound $d_2 \le r_2-s$ can be established in an analogous manner. This completes the proof of the converse in Theorem~\ref{thm:mimobcsdof-2}.

\section{Conclusion}
\label{sec:conclusion}

We study the achievable secrecy degrees of freedom
region for a two-receiver MIMO broadcast wiretap channel where only   the rank
(or an upper bound on the rank) of the eavesdropper channel matrix is known to the legitimate terminals.
While a direct extension of the single-user binning is sub-optimal, we show that the optimal degrees of freedom
can be obtained by 
simultaneously diagonalizing the  channel matrices of the legitimate receivers and carefully selecting
the transmission on the resulting sub-channels in order to share a common fictitious message between the two
receivers. We also extend the results to the case when an additional mutual-privacy 
constraint is imposed at the receivers.

While the present paper treats the case when the channels are static, extension to the case when the eavesdropper's channel
is time-varying, as considered in~\cite{xiang2010arb} in the single-user case, is left for future work.  It will be also interesting to extend our result to the case of more than two receivers. This could perhaps require finding a suitable extension of the GSVD transform to more than two channel matrices.
Another interesting future direction is to consider the case when the legitimate receiver's channel are also time-varying. 
One can also assume that the transmitter acquires  CSI with a unit delay. A connection to the recent result in~\cite{yang2011secrecy} appears natural and worth pursuing. Finally we note that this paper only considers the degrees of freedom of the broadcast network, which characterizes the pre-log of achievable rates. Such analysis is relevant if the network operates in the in the high SNR regime. The finer characterization of {\em constant-gap analysis} is left for future work .

\section*{Acknowledgements}
The authors thank the Associate Editor and the anonymous reviewers for several constructive suggestions that have improved the overall presentation of the paper. 

\bibliographystyle{IEEEtran}

\end{document}